\documentclass[11pt]{article}
\usepackage{amsmath,graphicx}

\numberwithin{equation}{section}
\def\eqn{Eq.~}
\def\eqns{Eqs.~}

\def\Ln{\mathop{\rm Ln}}
\begin{document}
%\doi{10.1080/1023619YYxxxxxxxx}
% \issn{1563-5120}
%\issnp{1023-6198} \jvol{00} \jnum{00} \jyear{2016} \jmonth{October}

%\markboth{Ronald E. Mickens}{Culling in Deterministic ODE
%Predator-Prey Models}

%%\articletype{MANUSCRIPT}

\title{\bf Consequences of Culling in Deterministic ODE 
Predator-Prey Models\thanks{Contact author: R. E. Mickens 
(rmickens@cau.edu)}}

\author{Ronald E. Mickens$^a$, Maxine Harlemon$^b$, Kale Oyedeji$^c$
\\\\
$^a$Department of Physics, Clark Atlanta University; Atlanta, GA 30314, USA
\\
$^b$Department of Biological Sciences, Clark Atlanta University;
\\
Atlanta, GA 30314, USA
\\
$^c$Department of Physics, Morehouse College; Atlanta, GA 30314 USA
}

\vspace{6pt} %%\received{v3.4 released June 2008}

\maketitle

\begin{abstract}
We show, within the context of the standard class of deterministic ODE 
predator-prey mathematical models, that predator culling does not
produce a long term decrease in the predator population.\bigskip


\bigskip
\end{abstract}

\section{Introduction}
Consider the interactions between predator, $y(t)$ and prey, 
$x(t)$, populations.  The time variable, $t$, is explicitly shown since
both populations are expected to change with time.

Now define culling as the external reduction of the predator population
at time, $t=t_0$, by an amount $y_c$, i.e.,
\begin{equation}\label{eq1.1}
y(t_0)\to y(t_0)-y_c,
\end{equation}
where
\begin{equation}\label{eq1.2}
0<y_c<y(t_0).
\end{equation}
Generally, this is done with the expectation that there will be a 
(long term) corresponding increase in the prey population.  
This action may be repeated into the future at either regular or irregular
time intervals.

The purpose of this work is to provide arguments which show that within the
context of standard mathematical predator-prey (P-P) models, based on
coupled ordinary differential equations, culling 
does not change the long term behavior/values
of either the predator or
prey populations.  This (almost) obvious and important consequence for
this class of models is generally not fully understood
and appreciated within the framework of population dynamic calculations.

Our interest in the topic of predator culling had its genesis in the
NEWS in Depth story, ``No proof that predator culls save livestock,
study claims," by Goldfarb \cite{2}.  He presented evidence which
indicates that the outcomes of predator culling may not be consistent
with prior notions of what should occur, namely, an end or at least a
reduction in predation.  Our conclusion is that the net effect of 
culling, again within the standard ODE P-P mathematical modelling
is zero, as stated in the previous paragraph.

The paper is organized as follows: Section~2 lists a number of the better
known P-P models.  This is done to provide an indication of the broad
range of mathematical structures which have been constructed for
P-P interactions.  Section~3 gives a summary of the generic 
properties expected of a P-P model and, in particular, the
important feature of the nontrivial fixed-point, i.e., the constant 
solutions
\begin{equation}\label{eq1.3}
x(t)=\bar x >0, \quad y(t)=\bar y>0,
\end{equation}
where the prey and predator can mutually exist.  Section~4 is
the center of the paper.  There, we discuss and derive the main
conclusion, i.e., culling does not lead to the desired outcome within
the frame work of deterministic ODE models.  We end, in Section~5, by 
briefly discussing other mathematical models for which the conclusions
of this paper may not hold.

\section{ODE Based P-P Models}
Below are listed five P-P mathematical models based on their formulation
in terms of coupled, nonlinear ODE's.  Note, we only consider the case
for which there is a single prey population, $x(t)$, and a single predator
population, $y(t)$.

The references \cite{4,5,7,8} provide background historical information
on these and other P-P models as well as giving, in several cases,
associated analytical and numerical analyses.  It should be noted that
the non-negative parameters, $(a,b,c,d,r,K,g)$ may, in different equations
have different physical units, as well as different interpretations as to 
their ecological meaning.

We now give a representative listing of P-P models which have been studied:

\centerline{\bf Lotka-Volterra}
\begin{equation}\label{eq2.1}
\frac{dx}{dt} = x(a-by),\quad
\frac{dy}{dt} = y(-c+dy),
\end{equation}

\centerline{\bf Verhulst-Pearl (Logistic)}
\begin{equation}\label{eq2.2}
\frac{dx}{dt} = rx(K-x)-bxy,\quad 
\frac{dy}{dt} = y(-c+dy).
\end{equation}

\centerline{\bf Gompertz}
\begin{equation}\label{eq2.3}
\frac{dx}{dt} = rx\Ln\left(\frac Kx\right) -bxy,\quad
\frac{dy}{dt} = y(-c+dy).
\end{equation}

\centerline{\bf Logistic-Ivlev}
\begin{subequations}\label{eq2.4}
\begin{align}
\frac{dx}{dt} &= rx(K-x)-y[1-\exp(-bx)],\label{eq2.4a}\\
\frac{dy}{dt} &= y\left\{ -c+d\left[1-\exp (-bx)\right]\right\}.
\label{eq2.4b}
\end{align}
\end{subequations}

\centerline{\bf Logistic-Holling-Leslie}
\begin{subequations}\label{eq2.5}
\begin{align}
\frac{dx}{dt} &= rx(K-x)-\frac{axy}{b+x},\label{eq2.5a}\\
\frac{dy}{dt} &= cy\left[ 1-\left(\frac y{gx}\right)\right]
\label{eq2.5b}
\end{align}
\end{subequations}

An examination of the book by May \cite{5} allows one to see how many 
other ODE based models can be constructed for P-P interactions.

\section{Generic Properties of P-P Models}
Realistic ODE based mathematical models of P-P interactions
should have certain features.  If we write
\begin{equation}\label{eq3.1}
\frac{dx}{dt} = xF(x,y),\quad\frac{dy}{dt} =yG(x,y),
\end{equation}
where $x(t)$ and $y(t)$ are the respective prey and predator 
populations, then in the $(x,y)$ phase-space \cite{6,7},
\eqns\eqref{eq3.1} must have (at least) three fixed-points
(FP), also known as constant or equilibrium solutions, to be a valid
model of a P-P system.  If we denote a FP by the notation
$(\bar x,\bar y)$, then a minimal PP model

(i) has a FP at ($\bar x^{(1)},\bar y^{(1)})=(0,0)$.  This
corresponds to a state for which there are no prey or predators.

(ii) has a FP at $(\bar x^{(2)}, \bar y^{(2)})=(\bar x,0)$.  For
this state, there are no predators, but the prey can exist by itself
at the equilibrium value $x(t)=\bar x$.

(iii) has a nontrivial FP at $(\bar x^{(3)}, \bar y^{(3)})=(x^*,y^*)$,
where the prey and predator can co-exist.

\begin{figure}[h!]
\centerline{\includegraphics[width=2truein]{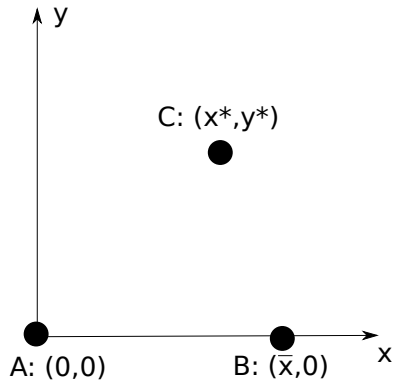}}
\caption{The location of the fixed-points (FP) for a generic
P-P system.  FP-A is unstable. FP-B is unstable.
FP-C may have neutral stability, i.e., it is a center, 
be a stable node, or be an unstable node.}
\label{fig1}
\end{figure}

These three cases are shown in Figure~\ref{fig1}.  However, detailed
investigation of these cases allows the following conclusions
to be reached \cite{3,4,5,7,8}:

(a) The FP's at $A=(0,0)$ and 
$B=(\bar x,0)$ are locally unstable.

(b) The FP at $C=(x^*,y^*)$ can have three types of stability \cite{6}:
neutral stability, corresponding to a center; a stable node,
or an unstable node.

\medskip
\hrule height.9pt

\hrule height.9pt
\medskip

\begin{figure}[h!]
\centerline{\includegraphics[width=3truein]{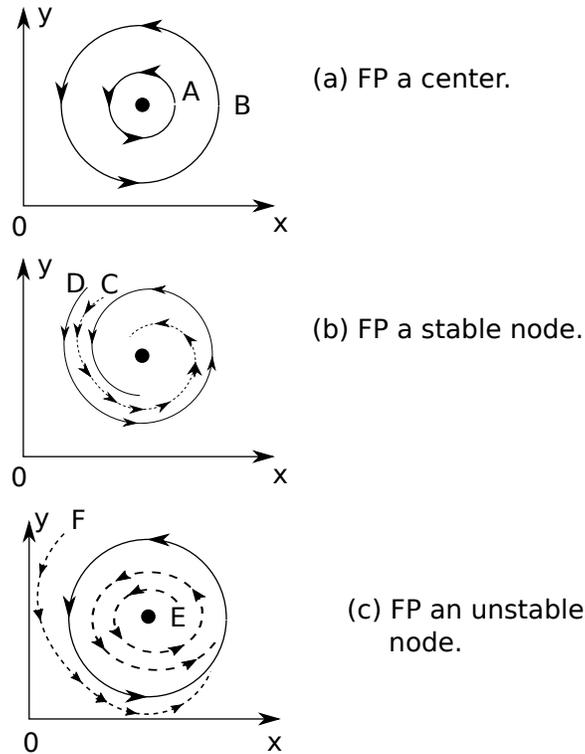}}
\caption{The heavy dot is the fixed-point (FP), located at
$(x^*,y^*)$. As indicated, there are three possibilities for the stability
of the FP.  In 2.c the continuous, closed curve is a limit-cycle.}
\label{fig2}
\end{figure}

Since our major interest is FP-C, we illustrate 
in Figure~\ref{fig2}, the outcomes of the three cases listed in (b).  
Note that the case for the center is not realistic since one
can select initial conditions, $x(0)=x_0>0$ and $y(0)=y_0 >0$, such 
that the predator and prey oscillation amplitudes can be arbitrarily
(but finite) large.  Consequently, this implies that the
Lotka-Volterra model, see \eqn\eqref{eq2.1}, while widely used to illustrate
P-P dynamics, is not realistic.  We will not consider this case any 
further in this paper.

Figure~\ref{fig2}-b, shows the situation for the co-existence state
being a stable node.  For this case all trajectories spiral into
the F-P $(x^*,y^*)$.

The third case is the most interesting; see Figure~\ref{fig2}-c.  This
illustrates that an unstable node may be surrounded by a stable limit-cycle
\cite{4,5,6,7,8}, and all initial conditions give rise to 
trajectories approaching the limit-cycle.  The maximum and minimum 
population values for both predator and prey, along with the period are 
determined by the parameters and not the initial conditions.

\section{Effects of Culling the Predator}

\begin{figure}[h!]
\centerline{\includegraphics[width=2truein]{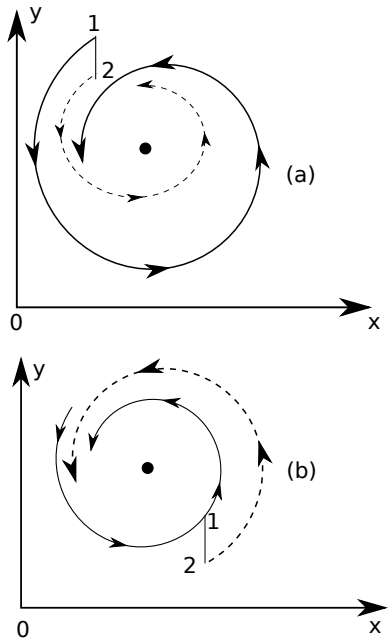}}
\caption{The FP, $(x^*,y^*)$, is a stable node.}
\label{fig3}
\end{figure}

\begin{figure}[h!]
\centerline{\includegraphics[width=2truein]{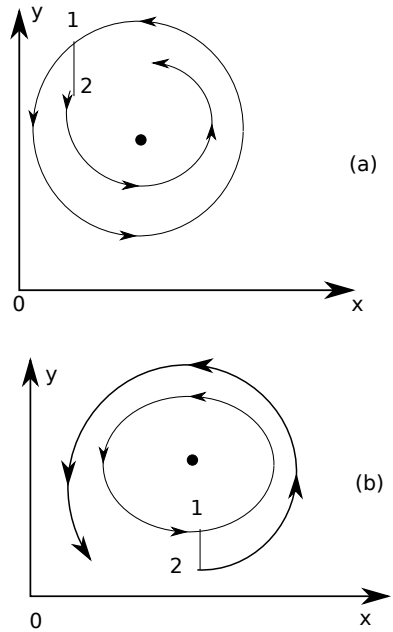}}
\caption{The FP, $(x^*,y^*)$ is an unstable node, inside a stable 
limit-cycle.}
\label{fig4}
\end{figure}

Culling of the predator, at time $t_0$, is equivalent to changing
the phase-plane point $(x(t_0),y(t_0))$ to $(x(t_0),y(t_0)-y_c)$, where
$y_c>0$ is the magnitude of the cull. Note that in
Figures~\ref{fig3} and \ref{fig4}, the ``1" denotes the location of
the trajectory at time $t_0$, i.e., $(x(t_0),y(t_0))$, while
``2" is the phase-space position after culling has taken
place, i.e., $(x(t_0),y(t_0)-y_c)$.

Figure~\ref{fig3} illustrates the fact that if the FP is a stable node, the
result of culling is that the system returns to the FP.  In other words,
culling does not change the long term behavior of the predator-prey
system.

Figure~\ref{fig4} shows the consequences of culling the predator for 
the case where the FP is an unstable node and it is surrounded by a stable
limit-cycle.  Again, the long time dynamics returns both the predator and
prey populations to the original limit-cycle.

\section{Summary}

The general conclusion is that the standard deterministic ODE based models
for a single predator and a single prey predict that culling of the
predator will not have any long term effects on the predator/prey
populations.  Thus, to the degree that actual data may give the opposite result,
it follows that for this situation, other types of mathematical
models are required.  One possibility is the consideration of
models which include stochastic effects \cite{1,5}.

\section*{Acknowledgement}
This work is an expansion of the discussion given in an e-letter
to SCIENCE, which comments on the article by Goldfarb \cite{2}.
This e-letter can be found at
{\tt http://science.sciencemag.org/content/353/6304/1080.e-letters}.


\begin{thebibliography}{999}
\bibitem{1} P. R. Ehrlich and L. C. Birch, Amer. Natur. {\bf 101} (1967),
pp.~97--107.

\bibitem{2} B. Goldfarb, SCIENCE {\bf 353} (2016), 1080--1081.

\bibitem{3} A. N. Kolomogorov, Giorn. Instituto Ital. Attuari {\bf 7}
(1936), pp.~74--80.


\bibitem{4} R. M. May, SCIENCE {\bf 177} (1972), 900--902.

\bibitem{5} R. M. May, {\it Stability and Complexity in Model 
Ecosystems} (Princeton University Press, Princeton and Oxford, 1974,
2nd edition), Chapter~4.

\bibitem{6} R. E. Mickens, {\it Mathematical Metehods for the Natural and
Engineering Sciences} (World Scientific, London, 2004), pp.~132--144.

\bibitem{7} J. D. Murray, {\it Mathematical Biology} (Springer-Verlag,
Berlin, 1989), pp.~63--78.

\bibitem{8} M. Rosenzweig, SCIENCE {\bf 171} (1971), 
385--387.

\end{thebibliography}
\end{document}